# 3 Metamodelling

## State of the Art and Research Challenges

Jonathan Sprinkle[1], Bernhard Rumpe[2], Hans Vangheluwe[3], and Gabor Karsai[4]

[1] University of Arizona, Tucson, AZ, USA
sprinkle@ECE.Arizona.Edu
[2] RWTH Aachen University, Germany
http://www.se-rwth.de
[3] McGill University, Montreal, Canada
hv@cs.mcgill.ca
[4] Vanderbilt University, Nashville, TN, USA
gabor.karsai@vanderbilt.edu

**Abstract.** This chapter discusses the current state of the art, and emerging research challenges, for metamodelling. In the state-of-the-art review on metamodelling, we review approaches, abstractions, and tools for metamodelling, evaluate them with respect to their expressivity, investigate what role(s) metamodels may play at run-time and how semantics can be assigned to metamodels and the domain-specific modeling languages they could define. In the emerging challenges section on metamodelling we highlight research issues regarding the management of complexity, consistency, and evolution of metamodels, and how the semantics of metamodels impacts each of these.

## 3.1 Metamodelling: State of the Art

Models are powerful tools to express the structure, behavior, and other properties in mathematics, each of the hard sciences and in all areas of engineering. While models are very common, an explicit definition of a modelling language and an explicit manipulation of its models is tightly connected to computer based tools. Additional power can be gained by explicit definition and computer based manipulation of models e.g. in CAD, control engineering, algebraic mathematics and of course computer science. To be able to manipulate models, their language needs to be specified as model of these models—metamodels. In this section, we describe the state of the art for metamodelling, including the metamodelling of data structures, as well as the metamodelling of languages systems where appropriate.

### 3.1.1 Concepts in Metamodelling

Metamodelling (literally, "beyond Modelling") is the Modelling of models. In their most common use, metamodels describe the permitted structure to which models must adhere [1]; although out of the scope of this chapter, meta-meta-models formally describe metamodels, as they define the core abstractions permitted in metamodelling. In fact, some metamodelling languages are self-descriptive [2, 3]. A metamodel therefore describes the syntax of the models [4]. Through various extension mechanisms and additional rules with this representation of the syntax of models metamodels can also help to define the semantics of models, as we discuss later. The layered approach to modelling (through metamodelling) is depicted in Figure 3.1.

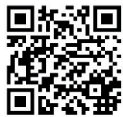

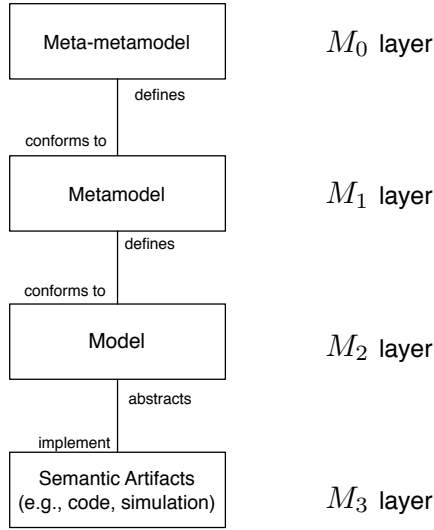

**Fig. 3.1.** The Four-layer metamodelling Architecture [5, 2, 6]. Some publications reverse the numbering of these layers, but we use this numbering, as platform transformations may create layers of arbitrary depth (e.g., $M_3$ may in fact be models defined in another layer $M_2$).

In the four-layer approach (generalized to $n$-layers in the MDA, discussed next) artifacts in each layer conform to, and/or are abstracted by the more abstract layer adjacent (in this case, the layer with a lower subscript). Thus, semantic artifacts in the $M_3$ layer are abstracted in models from the $M_2$ layer, which in turn conform to the metamodels from the $M_1$ layer. As these layers of abstraction are traversed, the role of each abstraction layer changes.

For example, a model is an encoding of some application or design in a different abstraction. Metamodels constrain the structure (and perhaps behavior) of models, but metamodels are relevant to all designs, not just a specific design. A widely-known example of this is that an XML document conforms to some type definition (either a DTD or XSD schema), but given only a schema, it is not possible to recover a particular XML file. When modelling languages, a metamodel basically needs to be considered a model of the abstract syntax of a language.

In Figure 3.1 we mention semantic artifacts. These are data, running programs, files, etc. that have some *meaning* in another context, e.g. by the user. They are artifacts in that they are produced through the design process. The running pro­gramm is regarded as a semantic artifact by a number of approaches, if it is gene­rated from high-level models, such as a state chart or dataflow model, while others regard code mainly as another (and final) syntactic representation of the system to be developed. For a metamodelling approach to have significant impact, some of those artifacts must eventually be produced in the design process; else, the mo­delling process is best classified as sophisticated documentation.

## Metamodelling: A Design Process

Generally, a design process that utilizes metamodelling first involves abstraction of the concepts of some domain or application into the appropriate meta-types (these are defined by the meta-metamodel $M_0$), using the metamodelling tools working on the $M_1$ layer. Metamodels can apply various archetypal concepts to constrain how models models are built, as shown in Table 3.1. The informed rea­der will see a dramatic similarity of these concepts with class modeling for soft­ware design. In fact, the visual representation chosen for many object-oriented modelling environments, and of metamodelling languages, is most commonly that of UML class diagrams.

**Table 3.1.** Archetypal abstractions used in metamodelling (adapted from [2, 4])

| Archetypal Concept | Description |
|---|---|
| Class | Specific classes of entities that exist in a given system or domain. Domain models are entities themselves and may contain other entities. Entities are instances of classes. Classes (thus entities) may have attributes. |
| Association | Binary and n-ary associations among classes (and entities). |
| Specialization | Binary association among classes denoting an IS-A relation. |
| Hierarchy | Binary association among classes denoting "aggregation through containment". Performs encapsulation and information hiding. |
| Constraint | An expression that defines the (statically computable) correctness of part of the model: only if all these constraints evaluate to true, the model is called "well formed". |

A visual depiction of the metamodelling design process can be seen in Fi­gure 3.2. In this figure, the metamodelling Interface corresponds to tools on the $M_1$ layer, and the Modeling Environment corresponds to tools on the $M_2$ layer, while the Application Domain corresponds to tools in the $M_3$ layer. As the appli­cation evolves, changes are made not to the $M_3$ layer, but to the more abstract layer ($M_2$). Similarly, as the Modeling Environment requires new types, they are modified in the model of the modeling environment (the Metamodel Specifica­tions). As we discuss in Section 3.1.5, metamodels that specify languages can also denote the concrete syntax for language elements, and constraints for the language, in the metamodel.

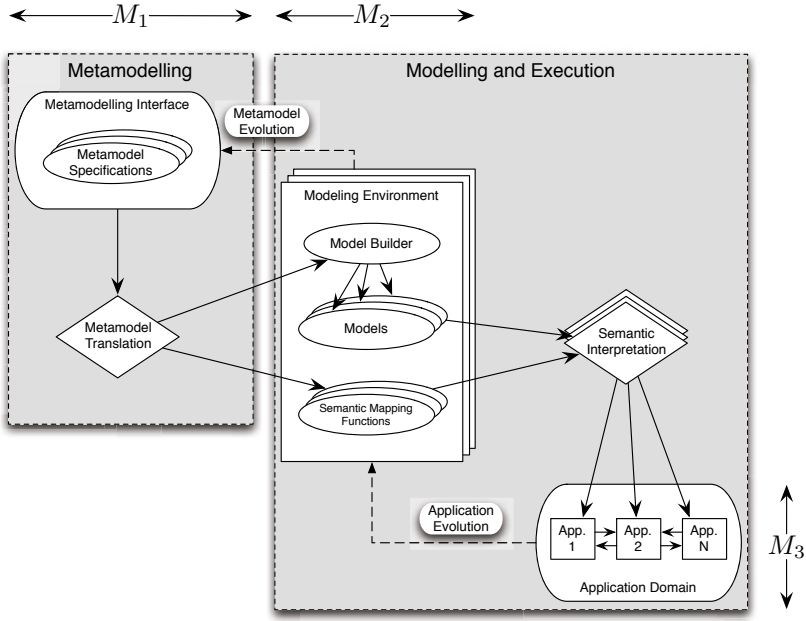

**Fig. 3.2.** The metamodelling design process

A concrete example may help to better understand the metamodeling design process. Let us consider the application domain ($M_3$) of electronic control units (ECUs) for automotive applications. The tool ($M_2$) should permit components to be connected to one another, and for components to be defined in terms of mathematical operations on their multiple input/output connections. There should also be constraints that prevent the outputs of two components to be connected to one another, for example. The metamodelling interface ($M_1$) can be used to create the tool, $M_2$, through the specification of the abstract syntax that permits these kinds of applications to be modeled. Metamodel translation synthesizes the tool, $M_2$, and semantic interpretation of models built using $M_2$ generates, for example, the embedded code for each component, a schedule for execution of components on a real-time operating system, logging functions for debugging purposes, and other necessary features in the application domain of ECUs.

In the event that a particular design ($M_3$) should be changed, the models created in $M_2$ should be modified, and then the semantic interpretation should be performed again. This is called *application evolution.* If the domain somehow changes, perhaps through additional constraints, new types that become available, or changes in design philosophy, then *metamodel evolution* must be performed, in order to change the design environment $M_2$. As shown in Figure 3.2, $M_2$ should be evolved by changing the metamodel specifications ($M_1$) and performing metamodel translation again.

**Archetypal Metamodelling Abstractions**

All major metamodelling approaches permit some significant subset of the basic abstractions shown in Table 3.1 [3, 7], though various tools may use a slightly different nomenclature [8]. As an example, the fundamental abstraction of "information hiding" is usually implemented using containment (as in hierarchical states).

Once the metamodels are defined in the design process, some transformation process generates the semantic artifacts necessary to continue in the design. This may be the generation of software skeletons that implement a class diagram, or the synthesis of configuration files that permit the use of a generic modeling tool. Textual modelers may use the metamodels to generate configurations for parsers and lexers to operate on text files that conform to the defined metamodels.

This synthesis process (metamodel translation, in Figure 3.2) maps types defined in the metamodel to concrete abstractions that an end-user will utilize to abstract their model-based design. Once the design of the application or system design is encoded in terms of meta-types (using tools from the $M_1$ layer), some transformation from the instances of these meta-types into the semantic domain is performed. In the remainder of this section, we will discuss some significant state-of-the-art approaches to metamodelling.

### 3.1.2   Meta Object Facility (MOF)

Central to the design and implementation of the UML 2.0 infrastructure and superstructure is the concept of model transformations between varying layers of abstraction. In order to permit these transformations (model to model, or as instances, object to object), some additional specification must be used to describe the structure of these objects—and this is termed the *Meta-Object Facility* (MOF).

The purpose of metamodelling through MOF is to describe the models in these various layers using common Modelling abstractions. This permits homogeneous access to models at all layers using reflection, standardizes access across tools through a common API, and permits serialization of models through the XMI standard. The specification of the MOF standard itself is well-described in the OMG document governing MOF (see especially [9]), and we do not attempt to fully describe those formalisms and terms here. However, we will describe the modelling concepts used by MOF to perform metamodelling, and we will do this from the perspective of the *Essential* MOF (EMOF), as described next.

### 3.1.3   Essential MOF (EMOF)

One major benefit of Modelling languages is to include concrete formalisms that makes modelling of particular concepts easy. For metamodelling, however, a significant amount of freedom in specification can lead to complexity when various models need to be updated. Thus, the use of an essential subset of a metamodelling language can insulate created models from changes to the metamodel.

This is the concept behind the Essential MOF (EMOF), which is a subset of the Complete MOF (CMOF) [9]. We present here the key concepts to EMOF, so that they can be compared to those of other metamodelling languages. In essence, what we present here is the metamodel of EMOF.

### Reflection, Identifiers and Extension

Basic assumptions for using EMOF (and CMOF) include the ability to utilize reflection, extension, and identifiers. *Reflection* is the ability of an object to determine its type (class, or "metaobject" in the EMOF vocabulary, and its associated metadata). *Extension* is the ability of an object to be dynamically annotated with `name-value` pairs. This permits some amount of runtime Modelling of particular objects, enabling an object to create new data fields which it could later use, without creating a new type. Note that when using extension, only that object receives these new `name-value` pairs, they should not propagate to all objects of that type. Finally, *identifiers* are a way for objects to maintain uniqueness regardless of any values with which it is instantiated or any extensions with which it is annotated.

In essence, two of these three concepts correspond to key attributes of Object-Oriented languages: an object is unique, an object knows its type. The other (extension) is a novel introduction to models when compared to textual languages, as in order to extend a class in a textual language requires creation of a new type.

### EMOF Classes

The fundamental metamodelling archetypes are easily visible in Figure 3.3 (compare to Table 3.1). However, as this is the MOF metamodel, such concepts are rewritten slightly. MOF permits `Class` objects (which inherit from *`Type`*). These `Class` objects will be able to contain `Property` and `Operation` features. In turn, the `Property` and `Operation` features that belong to a class are further made up of `Parameter` objects, or associated with `Property` objects.

It should be clear to the reader from examining the kinds of attributes in this metamodel that one major goal of MOF is Modelling *software models*. That is, the `Class` object has a specific data member `isAbstract`, which is an attribute commonly associated with software architecture. Not all metamodelling techniques are used simply as abstractions for software models, as we discuss in Section 3.1.5.

Given the wide acceptance of EMOF as a metamodelling framework, there are some key features and benefits to EMOF. It is possible to serialize EMOF models using the accompanying XMI standard (which provides mapping rules from EMOF to XML). There are also mappings from EMOF to Java, so as to generate software architectures and APIs from the models. Using these transforms, it is also possible to generate reflective operations in software, to permit manipulation of metamodel elements.

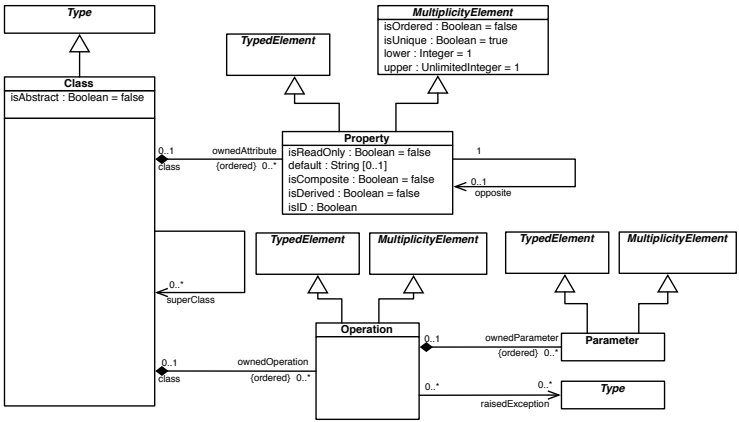

**Fig. 3.3.** The EMOF metamodel [10]

### 3.1.4   Eclipse Modelling Framework (EMF)

Similarly to EMOF, the Eclipse Modelling Framework (EMF) is a facility for
building models of data structures. EMF, as it is tied to a particular tool
(Eclipse), presents some additional benefits in that it can generate refined tools
and applications that are tailored for Eclipse. At root, it is still quite closely tied
to creating software models. EMF is a restricted subset of UML class diagram
concepts, namely the definitions of *classes*, *attributes* belonging to those classes,
and *relations* between classes.

Accompanying the EMF toolsuite is a set of plugins that permit reuse of EMF
models. Among the most significant are tools that permit editing EMF models
(and customizing EMF editors), and synthesizing software from EMF models.
These are discussed further in metamodelling-languages surveys, such as [11], as
well as the EMF documentation.

The popularity of the Eclipse toolsuite brings with it a plethora of Eclipse,
and EMF, plugins and tools that use the serialization that comes with the EMF
use of XMI standards, and the implicit tool interchange that is possible through
popular acceptance of the Eclipse platform.

There is a companion modeling framework for the visualization of EMF mo-
dels, called the Graphical Modeling Framework (GMF), which is part of the
Eclipse toolsuite. GMF uses the Graphical Editing Framework (GEF) in order
to interface with domain models graphically, and can leverage existing EMF me-
tamodels to bootstrap the visual language definition in GMF. Some features of
domain-specific modeling, such as constraint specification and multi-aspect vi-
sualization, are not yet part of the GMF toolsuite, but it is nonetheless a strong
tool for modeling of Java-based applications.

### 3.1.5   Metamodelling of Languages

The domain-specific modeling approach using metamodelling treats metamodels as language specifications, not software structure specifications. This difference from the common uses of EMOF and EMF distinguishes MetaGME [12], MetaEdit+ [13], and AToM$^3$[14, 15] tools. A common use of language generators is to synthesize domain-specific languages [16], and a rich legacy of application in this area can be found in the proceedings of [17, 18, 19, 20], and also in [21].

Language metamodelling defers representation and usage to a language or model editor. Meta-configurable editors such as GME, AToM$^3$, and MetaEdit+, are capable of reusing the same editor framework for many different languages. Eclipse's GEF also permits single editor multi-configuration reuse. However, in language use (alternatively, modeling environment use), the issues of concrete syntax and visualization must be addressed. We discuss this in Section 3.1.7.

An additional property found in the metamodelling of languages is the ability to specify selective visualization (also known as *aspects* or *viewpoints*). These properties permit filtering of the visualization space for an intuitive subset of the design, as partitioned at design time. An example of these properties is seen in Figure 3.4, where subsets of each object are visible, depending on the aspect in use. In this particular example from the signal processing domain, certain computational blocks may share parameters, but these blocks are not functionally connected. For the purposes of design it can be convenient to see what computational blocks share the same parameters, but if this information were shown in the same screen as the functional connections, it would be difficult to understand the diagram.

The final property we discuss with respect to language modelling not regularly found in data Modelling is that of constraint specification within the metamodel. Constraints may exist for certain data Modelling applications, but at the language level, such constraints can *prevent* or *restrict* the ability of a modeler to create certain constructs that are known *a priori* to have no well-defined semantic interpretation (or perhaps a disallowed, but known, semantic interpretation). These constraints may be specified in terms of the OCL (Object Constraint Language) [22].

A common use of constraints is to permit simple metamodel specifications, with small exclusions from their use. For example, a metamodel may define connections between ports of container objects (such as that shown in Figure 3.6a). However, a constraint can prevent the connection of two output ports to one another *unless* those output ports are at different levels of hierarchy (i.e., passing a value on to a parent's output port). Such a constraint can be written in OCL as:

```
OutPort.attachingConnections( BufferedConnection )->forAll( c |
    c.connectionPoints( "src" )->theOnly( ).target( ).parent( ).parent( ) =
  c.connectionPoints( "dst" )->theOnly( ).target( ).parent( ) )
```

This concisely states that if an `OutPort` object participates in an association of kind `BufferedConnection`, that the grandparent of the `src` must be the parent of the `dst`. This prevents two `OutPort` objects of the same `Component` from connecting through this kind of association. Constraints provide a powerful means to restrict the modeler from creating ill-formed models, while maintaining a language that is easy to compile.

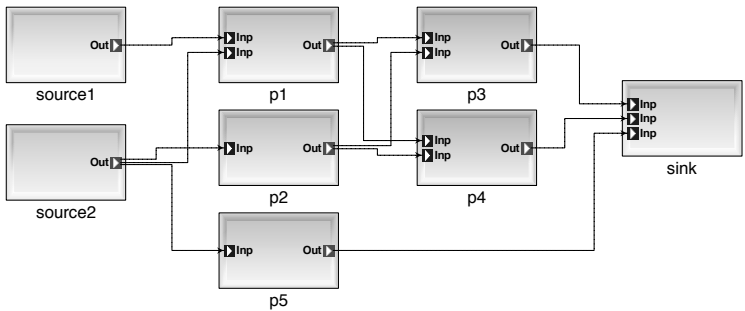

(a) Signal flow structure.

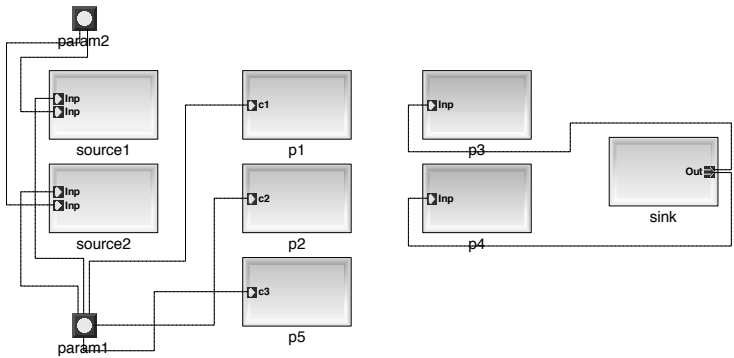

(b) Shared parameters of the system.

**Fig. 3.4.** These two figures show how the same structural elements can be shown in different aspects to visualize elements of the system effectively. In (a), the structure as related to the signal flow is given. In (b) the parameters shared between components are easily seen—and easily changed.

### 3.1.6    Textual Metamodelling

Metamodelling is useful in textual, as well as graphical/visual, languages. In fact, there are certainly cases where a textual language is preferred [23]. When considering the traditional methods of specifying grammars (e.g., Backus Naur Form [24], and Extended BNF), it is apparent that such specifications do define the abstract syntax of a language. Tools that generate parsers and lexers for such grammars (such as antlr, bison, etc.) are the textual analogs to the abstract syntax tree generators found in modeling environments [25]. The application of programming language types to their semantics is well-studied [26], and rigorous treatment of their specification can permit subtle understandings.

### 3.1.7    Concrete and Abstract Syntax

The differences between concrete and abstract syntax are well known, and well-studied [26, 27]. However, their application to new modelling languages brings into question how to specify concrete syntax best during the metamodelling phase of language design, as well as how the abstract and concrete syntaxes are used.

For textual modelling languages, some concrete syntax is required in order to streamline model construction. Although visual language developers have been among the most vocal proponents of modelling, there is significant research in textual domain-specific languages, because of their better efficiency, both in language definition and use [28].

For graphical modelling languages, a (default) concrete syntax can be synthesized directly from the metamodel, as long as a default concrete syntax is provided for each archetypal type. The GME tool e.g. has generic types as defined by the meta-metamodel, which provide a default concrete syntax if not overridden at the metamodelling level. Overriding the concrete syntax is fairly straightforward (many tools such as Simulink and LabVIEW permit this as well), it can be done at the $M_2$ or $M_1$ level. There are important questions that must be resolved with regards to the semantics of concrete syntax changes at any level, especially for tools that are domain-specific in nature, and depend on an intuitive understanding of visual models; we discuss these issues below. Other tools such as DiaGen [29] utilize concepts similar to that of GME to attach visualization attributes to the nodes and edges graph that encodes the model.

#### Metamodelling Level Concrete Syntax Specification.

Defining a specific concrete syntax is possible for graphical languages by specifying a glyph or glyph-generator that will provide a visualization (perhaps context-specific) for a particular type in the language. Then, for every instance of this type that is visualized, this image (or the imaged produced by the glyph-generator) replaces the default value. This is very useful for simple domain-specific visual languages, where concrete domain items can be composed easily with other domain items.

#### Modelling Level Concrete Syntax Specification.

Redefining the appearance of a model, namely the concrete syntax, at the modelling level is also possible, though not as widespread as overriding at the metamodelling level. Whereas redefinition at the metamodelling level operate for each created instance, redefinition at the modelling level overrides just for one particular instance thus allowing allows individual shapes for each model element. Such overrides are to some extent also questionable due to the fact that (for some reason) the metamodel designer chose a different concrete syntax. Why is this concrete syntax being overridden? Will this confuse other modelers using this model? Any confusion in these areas will reduce the positive impact seen in the utilization of modelling languages to specify a design, as new users will be unable to distinguish between semantics of the language, and visual preferences of another modeler.

**Concrete Syntax**

Concrete syntax is carefully chosen to represent domain concepts (for domain-specific languages), as "syntactic sugar" (for DSLs as well as general-purpose languages), or to otherwise make programming or Modelling easier. For the semantic interpretation of a language, however, there is the possibility that the concrete syntax *could* be used in semantics definitions, or that different variants of concrete syntax make the visual representation of a model ambiguous to developers. The inability to distinguish ambiguous representation in a screenshot of the model is shown in Figure 3.5.

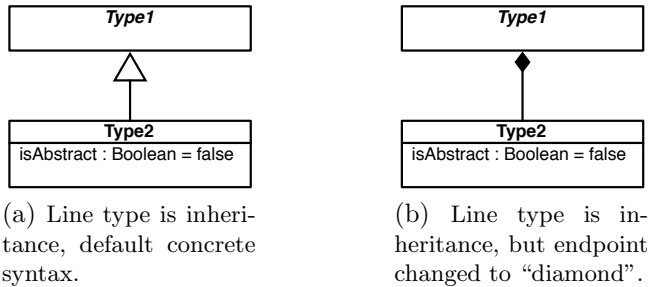

(a) Line type is inheritance, default concrete syntax.

(b) Line type is inheritance, but endpoint changed to "diamond".

**Fig. 3.5.** Changing the concrete syntax at modelling time can lead to confusion. In (a) the default concrete syntax for inheritance is used. In (b) a modeler has changed the appearance of the line, but the semantic interpretation will still be inheritance.

It is a good practice, generally, to *not* use the concrete syntax details in the mapping of semantics, but to depend entirely on the abstract syntax tree. An interesting research challenge would be general-purpose tools that could identify issues such as these in completed models (or perhaps in their semantic mapping) as potential design flaws in the model, language, or language compiler.

### 3.1.8   Type System

Type systems in traditional programming languages are established at language-design time. In (typed) programming languages as well as in math, a type is basically a description for a set of values together with a set of operations to manipulate these values. In the metamodelling setting this approach needs to be adapted to meta-type structures. Different to the programming language approach, metamodelling approaches tend to merge typing and the meta-level. The challenge of defining a type within the model while applying it at the same time can then be met through the use prototypes and clones as discussed below.

For modelling language types the most common mode of type definition is through specification in a metamodel (as described in Section 3.1.1). Using this mode of definition, the traditional object-oriented abstractions of type definitions can be leveraged into the modelling language. These include notions of inheritance, containment, and association. Propagation of model features (i.e., subclasses have all features of parent classes) through generalization/ specialization relationships in the metamodel provide a means to effect polymorphic behaviors at model execution or interpretation time. Please note that it is rather convenient to lift the type infrastructure of the object-oriented realization of the metamodel into the defined modelling language. However, one could build an entirely different type system, and when the oo style of typing doesn't fit we are even forced to do so.

There are the following phases of type system use and specification:

– Meta-metamodelling time: specification of the fundamental meta-types, which define how types permit containment, association, attribute values, etc.
– metamodelling time: specification of the metamodel, using meta-metamodel types, in order to define model types (e.g., domain-specific concepts), and the abstract syntax of the language for language-generating metamodels.
– Modelling time: specification of certain clones and clone structures as templates for further instantiation.

New patterns and structures not conceived at language-design time, however, may emerge *after* the metamodels have been designed. Many modelling environments permit the Modelling-time specification of new type systems, which permits a modeler to develop a new "type" out of composition and association of instances of domain types. In this case, the new "type" may be reused, re-instantiated multiple times, and may propagate changes made to the type to any instances of that type.

In order to distinguish easily between types (defined at metamodelling time), and Modelling-time types, we use the following nomenclature (from [30]):

– **prototypes**: modelling-time types; and
– **clones**: instantiations of prototypes.

We explore this modelling-time types behavior through an example.

## Types and Clones at Modelling Time

Consider the language defined by the metamodel shown in Figure 3.6a[1], and models built using this language shown in Figure 3.6b. Now, let us consider that model C2, contained in Component1 is a *clone* of the *prototype* Component2. This would mean that for each object contained in Component2, there would be a corresponding object of the same type (and participating in corresponding *internal* associations) in C2. By *internal* associations, we mean to say that the association is contained by that model (and is not an association that resides outside the type).

Regarding the attribute values of these models, whenever a clone of a prototype is created, it receives the attribute values of the prototype. From this time on, there are several semantic issues which must be addressed by the modelling environment.

---

[1] This metamodel is reused in Chapter 9 in order to discuss the evolution of models.

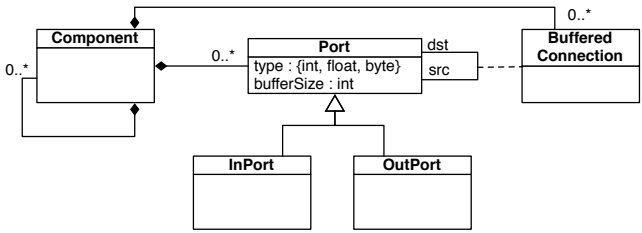

(a) The metamodel allows objects of kind `Port`, which is specialized as `InPort` and `OutPort`.

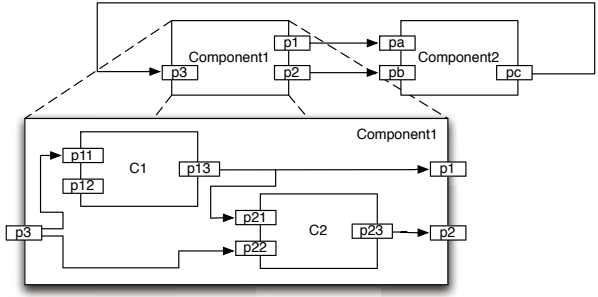

(b) A model built using the metamodel in (a). The contents of `Component1` are shown to display the additional associations in which its `Port` objects play a role.

**Fig. 3.6.** (a) A metamodel allowing port interconnection between components. (b) A model built using the metamodel in (a). The "arrow" end of the connections represent the `dst` role.

(1) Are attribute values of the clones permitted to be modified?
(2) If an attribute value of a clone is modified, and the prototype attribute value is modified, what then should be done for the clone model's attribute values?
(3) If the attribute values of the prototype change, should unmodified attributes of the clones be updated?

Tools and environments that permit prototypes and clones adopt a fairly consistent view of these questions. Both GME [12] and Ptolemy II [31] permit attribute value modifications of clones. In the event of changes to the original prototype, an attribute-specific copy-on-write behavior is utilized, where unmodified attribute values reflect the prototype values, rather than maintaining the values at instantiation-time. We discuss in the next section how selective permission to contain new objects in clones and prototypes can create some confusion.

It is generally up to the tool developer to determine how to visually depict prototypes and clones. If a separate browser that permits searching for or displaying only prototype and clone hierarchies is given, it is not necessary to even have a visual cue that a particular object is a clone.

A final restriction on clone models is that they cannot contain objects that are not contained in the prototype object (i.e., the correspondence function is bijective). Similar to attribute propagation, then, new models created inside an prototype propagate to *all* clones of that prototype. If an object contained within an prototype is deleted, all clones remove their corresponding object (and any associations to that model, as appropriate).

**Prototypes and Subprototypes at Modelling Time**
Subprototypes have a subset of the restrictions and constraints of clones. As a class diagram permits subclasses to specialize the structure of their superclass, a subprototype can add to the features of an prototype at modelling time. Thus, the restriction that there does not exist any object in the instance that does not correspond to a type-contained object is not necessary (i.e., the correspondence is injective from prototype to subprototype).

### 3.1.9    Merging of Metamodels

Given the ability of models to apply hierarchy and refinement as abstractions, and the fact that metamodels are models themselves, the ability to merge metamodels (others say "compose metamodels" structurally) is somewhat trivial. The semantics of this merge, however, deserves some discussion.

Consider two metamodels, $\mathcal{M}_1, \mathcal{M}_2$, such that $\mathcal{M}_1 \cap \mathcal{M}_2 = \emptyset$ (i.e. they do not share any meta-class). Now, consider that some elements from each of these metamodels can be related in a new, merged, metamodel, $\mathcal{M}_3 = \langle \mathcal{M}_1 \cup \mathcal{M}_2, f \rangle$. When merging the metamodels, some elements from each of the two metamodels must somehow be associated with one another. We can use this function $f$ to define appropriate relations between metamodel elements.

These relations can be considered as mappings for identity, or new properties. As discussed in [30, 32], the identity equivalence maps two types (one from each metamodel) as identical, and thus permits the associations and attribute values of those types to be a union of the definition in the two metamodels. More subtle is the desire to transfer only some of the associations and attribute values of a certain type. These are created as new metamodel types (found only in $\mathcal{M}_3$) which can inherit either the interface of the existing types, or the implementation of the existing types (meaning that containment and other relations are, or are not, transferred). For these subtleties, we refer the reader to [30, 32] for a full explanation with examples.

A short example of merging metamodels is given in Figure 3.7. In Figure 3.7a we see a simple modeling language for discrete systems. In this language, the behavior of the system is obtained by firing the `Behavior` model(s) in the current state. A simple modeling language for continuous time systems is shown in Figure 3.7b, with the capability to assign values to `Variable` objects through algebraic and differential (`Flow`) equations. In order to create a new language,

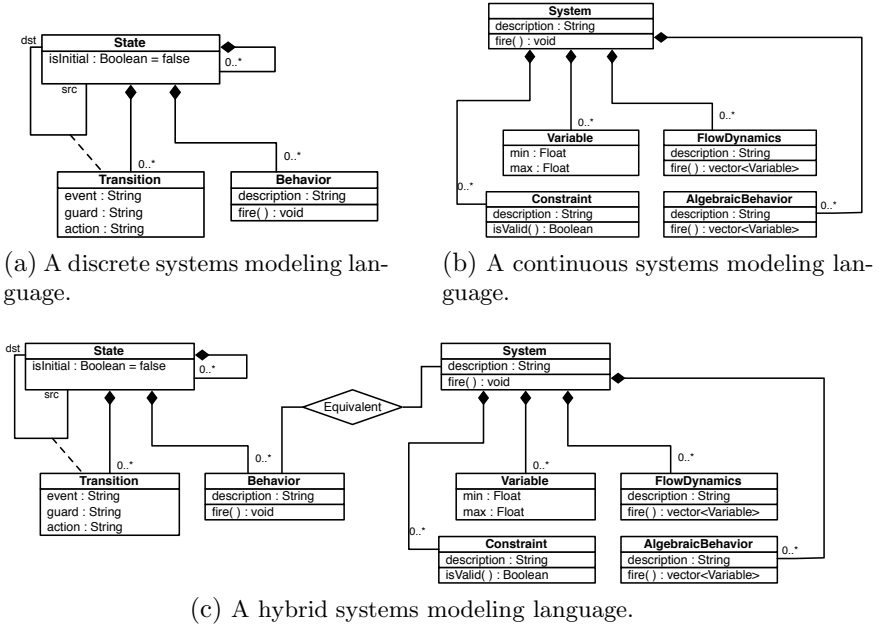

(a) A discrete systems modeling language.

(b) A continuous systems modeling language.

(c) A hybrid systems modeling language.

**Fig. 3.7.** Elements from the discrete, and continuous, domains are merged/composed in a new domain, and an equivalence relationship is used to indicate identity of one element in each metamodel

capable of modeling hybrid systems (those systems where each discrete state has a continuous dynamics), we can merge the two metamodels, and indicate an equivalence relationship between `Behavior` in the discrete systems language, and `System` in the continuous systems language. In this new modeling language, it is then possible to create new objects of kind `System` inside of a `State`, even though this is not explicitly shown through containment relations between `System` and `State`. It is possible to further assign relations between objects, using containment, association, or other relations.

An additional, metamodelling, concern is the propagation of constraints when metamodels are merged. These, and especially the issues of semantics, are research issues, and discussed in the following Section 3.2.

## 3.2    Metamodelling: Research Challenges

Metamodelling as a technology provides significant power to designers and users, and it has been thoroughly explored in terms of modeling data, software, and languages. Although many of the properties, semantics, and uses of metamodelling are now "solved" problems, there are significant research challenges still outstanding, regarding usability, evolution, intuitive representation, etc. We discuss these research challenges (in brief) in this section.

**A Unifying Issue: Semantics**

A unifying characteristic of nearly all outstanding research issues we discuss with respect to metamodelling is the issue of semantics. The meaning of composed, cloned, evolved, etc., models and metamodels may be unclear, depending upon the circumstances under which these operations are performed.

### 3.2.1     Semantic Attachment

We recall that metamodels are basically pure syntactic representations of the models they describe [4]. Significant strides have been made in attaching additional information to these metamodels, which in many approaches is called "semantic attachments". The significant issue in semantic attachment is not "can it be done?" but rather "what methods are appropriately efficient and intuitive?" A traditional compiler that traverses an abstract syntax tree to produce artifacts in the semantic domain can be readily produced (and tools to significantly automate the parsing and traversal have been developed [33, 28]).

Methods to ground semantics between metamodels to a common semantic domain show promise [34, 35] or through explicit definition of the semantics domain [36, 37, 38]. Utilizing those techniques, lossless, bijective, semantically-correct mappings between a metamodel and a semantic anchor such as the abstract state machine language (ASML) could foster semantically-correct interchange between tools, or from one semantic domain to another. One issue that deserves further research is the specifications of these mappings for complex metamodels, and their intuitive representation.

The expected use cases for attaching semantics to metamodels include:

– for the purpose of documentation/precise definition;
– facilitation of automatic verification of some property; and
– automated translation between tools.

### 3.2.2     Inference between Metamodels

As described in Section 3.1.9 it is possible to merge metamodels into a new metamodel, and (by design) mark certain metamodel elements as equivalent. The automation of this identification between two related (but separately specified) metamodels is an interesting research challenge. Issues are are present include:

– Semantic equivalence of inferred equivalent types in each metamodel;
– Visualization issues; and
– Propagation of constraints.

Among these, the semantic equivalence may require user interaction to determine. The propagation of constraints, however, presents a few interesting issues.

It may, for example, be possible to evaluate all models to determine whether constraints are violated *prior* to performing type inference. Perhaps, then, type inference is predicated on constraint satisfaction. On the other hand, issues such as selective propagation of containment (or containee) relationships may enable constraint satisfaction, so an intelligent approach to utilizing implementation and interface inheritance may permit some equivalence inference, while not violating any of the (union) of constraints.

### 3.2.3   Evolution of Models Driven by Metamodel Evolution

This issue presents tremendous challenge in the preservation of structure, constraints and semantics. As the metamodel evolves, e.g. because the tools are updated to a new version, it may be that models built using the metamodel will no longer conform to the evolved metamodel. In this case, evolution of the models (to conform to the new metamodel) may be required. This issue has been studied for visual languages [39], but further research is necessary to determine the best way to intuitively (and accurately) portray such evolutionary transformations. An interesting extension is the automation of such transformations *based on* changes to the metamodel in its evolution. More discussion is devoted to this complex topic in the Chapter of Model Management.

## 3.3   Conclusions

Despite the many tools available for metamodelling, the underlying mechanism of object-orientation has fostered that most of the metamodelling tools use a common set of abstractions with only slight variations. Using these abstractions, it is possible to raise the specification of a language and its tooling far above the implementation layer. Additional capabilities increase the power of metamodelling by permitting the synthesis of languages as well as automated or semi-automated analysis and synthesis techniques. Run-time modelling tools permit users to define their own prototypes, and leverage new patterns not anticipated at metamodel-design time, and the visualization of models can be carefully specified to ensure that information is appropriate presented to modelers. Multiple metamodels can be merged to specify new languages that appropriately integrate the concepts into one big metamodel .

Model-based engineering and in particular modelling of embedded systems benefits heavily from metamodelling due to the structure that metamodelling gives to models, and the semantics that can be attached to metamodels. Given this structure, the specification of semantics is easier, correspondences between metamodels can be denoted, parsers/lexers can be synthesized, and constraints can be evaluated. These capabilities are the foundations for raising the level of specification of systems to models, rather than low-level implementation.